\documentclass[journal]{IEEEtran}

\usepackage{amsmath, amssymb, bm, cite, epsfig, psfrag}
\usepackage{epstopdf}
\usepackage{graphicx}
\usepackage{algorithm,algpseudocode}
\usepackage{array}
\usepackage[margin=10pt,font=small]{caption}
\usepackage{bbm}
\usepackage{multirow}
\usepackage[usenames,dvipsnames]{xcolor}

\usepackage{subfigure}
\usepackage{etoolbox}
\usepackage{pbox}
\usepackage[width=0.48\textwidth]{caption}
\usepackage{colortbl}
\usepackage{fancyhdr}
\newtoggle{conference}
\togglefalse{conference} 
\graphicspath{{figures/}}

\def\beq{\begin{equation}}
\def\eeq{\end{equation}}
\def\beqa{\begin{eqnarray}}
\def\eeqa{\end{eqnarray}}
\def\beqan{\begin{eqnarray*}}
\def\eeqan{\end{eqnarray*}}

\def\tm1{t\! - \! 1}
\def\tp1{t\! + \! 1}

\usepackage{lipsum}
\usepackage{fancyhdr}
\usepackage{datetime}

\usepackage{multicol}

\usepackage{tikz}
\usetikzlibrary{calc}

\begin{document}

\newcommand\blfootnote[1]{%
  \begingroup
  \renewcommand\thefootnote{}\footnote{#1}%
  \addtocounter{footnote}{-1}%
  \endgroup
}

\pagestyle{empty}

\title{28 GHz Millimeter-Wave Ultrawideband Small-Scale Fading Models in Wireless Channels}

\author{
	\IEEEauthorblockN{Mathew K. Samimi, George R. MacCartney, Jr., Shu Sun, and Theodore S. Rappaport\\ NYU WIRELESS, NYU Tandon School of Engineering\\
mks@nyu.edu, gmac@nyu.edu, ss7152@nyu.edu, tsr@nyu.edu}\\

\thanks{The authors wish to thank the NYU WIRELESS Industrial Affiliates for their support, and S. Deng, T. Wu, M. Zhang, J. P. Ryan,  A. Rajarshi for their contribution to this project, and Prof. S. Rangan for valuable discussions. This work is supported by National Science Foundation (NSF) Grants (1302336, 1320472, and 1555332).}

}

\maketitle

\begin{tikzpicture} [remember picture, overlay]
\node at ($(current page.north) + (0,-0.25in)$) {M. K. Samimi et al., ``28 GHz Millimeter-Wave Ultrawideband Small-Scale Fading Models in Wireless Channels,''};
\node at ($(current page.north) + (0,-0.4in)$) {\textit{to be published in the 2016 IEEE Vehicular Technology Conference (VTC2016-Spring)}, 15-18 May, 2016.};
\end{tikzpicture}

\begin{abstract}
This paper presents small-scale fading measurements for 28 GHz outdoor millimeter-wave ultrawideband channels using directional horn antennas at the transmitter and receiver. Power delay profiles were measured at half-wavelength spatial increments over a local area (33 wavelengths) on a linear track in two orthogonal receiver directions in a typical base-to-mobile scenario with fixed transmitter and receiver antenna beam pointing directions. The voltage path amplitudes are shown to follow a Rician distribution, with $\bm{K}$-factor ranging from \text{9 - 15 dB} and \text{5 - 8 dB} in line of sight (LOS) and non-line of sight (NLOS) for a vertical-to-vertical co-polarized antenna scenario, respectively, and from 3 - 7 dB in both LOS and NLOS vertical-to-horizontal cross-polarized antenna scenario. The average spatial autocorrelation functions of individual multipath components reveal that signal amplitudes reach a correlation of 0 after 2 and 5 wavelengths in LOS and NLOS co-polarized V-V antenna scenarios. The models provided are useful for recreating path gain statistics of millimeter-wave wideband channel impulse responses over local areas, for the study of multi-element antenna simulations and channel estimation algorithms.
\end{abstract}

\section{Introduction}

Millimeter-waves (mmWave) presently constitute a key-enabling technology that will deliver multi-gigabits per second data rates for backhaul and mobile applications~\cite{Rap15,Khan11}. Many recent propagation measurements have demonstrated the viability of mmWave communications in both indoor and outdoor urban environments with the use of high-gain directional and steerable horn antennas~\cite{Rap13:2,Rap15_3,Haneda14,Hur15}, necessary to overcome the magnitude increase in free space path loss over conventional Ultra-High Frequency (UHF) and Microwave frequency bands, with little impact from rain or atmospheric attenuations over a few hundred meters~\cite{Rap15}. The measurement data contributed to omnidirectional and directional distance-dependent large-scale path loss~\cite{Rap15_3} and statistical channel impulse response models~\cite{Samimi15,Samimi15_3,Haneda14,Hur15}. Preliminary multiple-input multiple-output (MIMO) simulations using a measurement-based mmWave statistical spatial channel model (SSCM)~\cite{Samimi14} were carried out showing a large increase in wireless system throughputs~\cite{Sun14} over current Long Term Evolution (LTE) technology.

While large-scale path loss model parameters have been extensively published~\cite{Rap13:2,Rap15_3,Haneda14,Hur15}, the statistics of small-scale fading at mmWave frequencies have thus far received little attention, yet are vital when estimating path amplitude gains over a local area in MIMO simulations. Small-scale fading measurements over fractions of wavelengths in the UHF and Microwave bands in both indoor and outdoor environments have previously provided invaluable insight into spatial and temporal fading of multipath amplitudes, commonly characterized with a Rician distribution in LOS environments where a dominant path is present, and in NLOS channels with Rayleigh or lognormal distributions~\cite{Bultitude87,Rap91,Hashemi95,Karttunen98}. Such measurements also enable the study of autocorrelation properties of individual and total multipath signal amplitudes, equalization algorithms, and antenna diversity schemes to estimate signal fades, known to significantly impact system performance. 

Today's 3GPP, WINNER II, and COST geometry-based stochastic spatial channel models (SCMs), use Rician and Rayleigh distributions to recreate the small-scale fading statistics of path amplitudes in LOS and NLOS channels, respectively~\cite{3GPP:1,WinnerII,Calcev07:1,COST2100}. These SCMs were developed based on \text{1 - 6 GHz} RF propagation measurements, with up to 100 MHz RF bandwidth, that provided empirical evidence for Rayleigh-distributed path amplitudes~\cite{Kivinen01} in NLOS indoor channels. In the 3GPP SCM framework, a propagation path is assigned small-scale parameters, such as path delays and powers, generated from measurement-based statistical distributions. Each path is then further subdivided into 20 equal power subpaths, whose angle of departures (AOD) and angle of arrivals (AOA) are slightly offset from the small-scale path AOD and AOA. Small-scale spatial fading is synthesized by adding the 20 subpath (voltage) amplitudes, each subject to Doppler frequency shifts~\cite{3GPP:1}. Small-scale spatial channel models are essential for developing channel estimation algorithms of Doppler frequencies, path gains, and path delays of multipath components when evaluating modulation and coding schemes.

In this paper, the statistics of mmWave outdoor small-scale fading are obtained from 28 GHz measurements over a local area using a broadband sliding correlator channel sounder and a pair of high-gain directional horn antennas. Simple small-scale spatial fading models for individual multipath (voltage) amplitudes are extracted for line of sight (LOS) and non-line of sight (NLOS) environments, for both co- and cross-polarization scenarios. These models may easily be implemented in channel emulators that recreate channel impulse responses and realistic narrowband fading amplitude envelopes for short, sub-wavelength distances in multi-element antenna simulations~\cite{Tranter04,Fung93,Rap91_2,Samimi16}.

\section{28 GHz Small-Scale Fading Measurements}

The 28 GHz street-canyon small-scale fading measurements were performed on the NYU Brooklyn campus, using a 400 megachips-per-second broadband sliding correlator channel sounder, whose superheterodyne architecture is outlined in~\cite{Rap15_3}. The transmitter (TX) and receiver (RX) each used 15 dBi (28.8$^{\circ}$ and 30$^{\circ}$ half-power beamwidths in azimuth and elevation, respectively) directional horn antennas, with distances from the TX antenna to center of the RX local area ranging from 8 m to 12.9 m, with maximum TX power of 27 dBm. The largest measurable path loss was approximately 157 dB, with a 2.5 ns multipath time resolution (800 MHz RF null-to-null bandwidth). Directional antennas were utilized to emulate future mmWave systems that will consider electrically-steered multi-element antenna arrays and beamforming algorithms to enable high antenna gains at the TX and RX. The models presented herein are valid after beam searching is performed while the mobile device is in motion.  

\begin{figure}[t]
    \centering
 \includegraphics[width=3.5in]{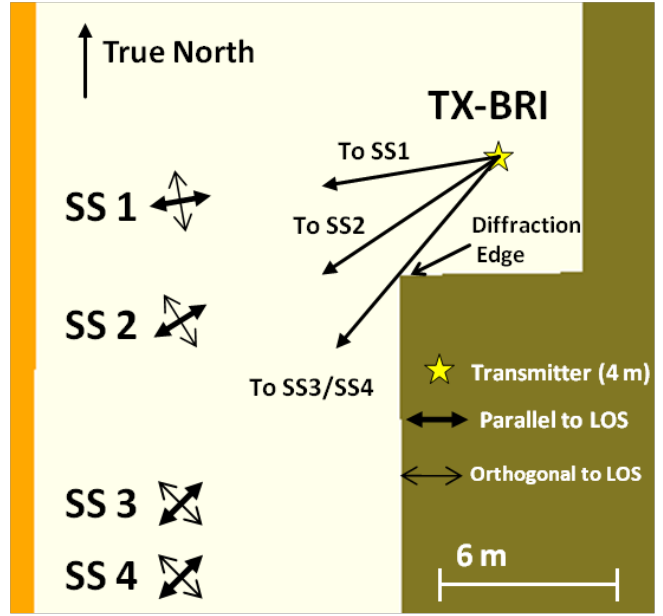}
    \caption{Map of NYU Brooklyn campus, where the 28 GHz small-scale fading outdoor measurements were conducted in a street canyon scenario, with one TX location (BRI), and four RX locations. The black arrows originating from the TX show the azimuth AODs used for each RX local area, while the orthogonal arrows represent the two axes of a cross over which a power measurement was made every 5.35 mm to simulate a virtual array of directional antennas.}
    \label{fig:map}
\end{figure}

The ultrawideband measurements investigated spatial and temporal fading and autocorrelations of the received multipath signal amplitudes over a local area for vertical-to-vertical (V-V) and vertical-to-horizontal (V-H) antenna polarization scenarios. One TX location, Bridge (BRI), and four RX locations were selected to conduct the measurements in LOS, NLOS, and a transitional LOS-to-NLOS environment to gain insight into the statistics of multipath component amplitudes in realistic mobile environments, as depicted in Fig.~\ref{fig:map}. At each location, the RX antenna was moved over a stationary 35.31-cm spatial linear track (33 wavelengths) in steps of $\lambda/2=$ 5.35 mm, and for each track position a power delay profile (PDP) measurement was acquired. Each PDP capture was an average of 20 consecutive PDPs, with a total PDP capture time of 818.8 ms. 

\begin{figure}[t]
    \begin{center}
        \includegraphics[width=3.5in]{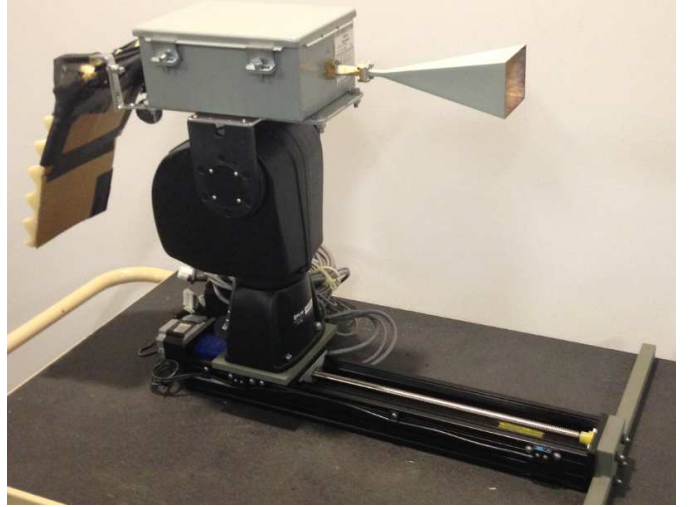}
    \end{center}
    \caption{The small-scale linear track used to study small-scale signal fading over fractions of wavelengths.}
    \vspace{-0.4cm}
    \label{fig:LinearTrack}
    \end{figure}

The TX and RX antennas remained fixed in azimuth and elevation during the measurement captures, and were positioned 4 m and 1.4 m above ground level, respectively, well below surrounding rooftop heights of approximately 40 m. In total, 66 PDPs were acquired over one track length measurement. The linear track was positioned such that the RX antenna could be spatially incremented towards the TX antenna, and laterally from left to right of the static TX beam, so as to capture the channel fading over two orthogonal receiver directions, as illustrated with arrows in Fig.~\ref{fig:map}. The spatial track laid entirely within the 3-dB beamwidth of the TX antenna for the two LOS locations to minimize the effects of antenna beamwidth, thus removing the need to de-embed the antenna patterns from the measurements. Fig.~\ref{fig:LinearTrack} shows a photo of the linear track used during the measurements.

\section{Channel Impulse Response Model}

The propagation channel is commonly described by the superposition of multiple traveling waves, where each impinging wave at the receiver is described by a complex (voltage) path amplitude, a delay, an azimuth and elevation AOD, and an azimuth and elevation AOA. The double-directional time-invariant complex baseband channel impulse response (CIR) $h_{omni}(t,\overrightarrow{\mathbf{\Theta}},\overrightarrow{\mathbf{\Phi}})$ can be expressed as~\cite{Samimi15,Steinbauer01},
\begin{equation}\label{eq1}
\begin{split}
h_{omni}(t,\overrightarrow{\mathbf{\Theta}},\overrightarrow{\mathbf{\Phi}}) &= \sum_{k=1}^N a_k e^{j\theta_k} \delta(t - \tau_k)\\
& \cdot \delta(\overrightarrow{\mathbf{\Theta}}-\overrightarrow{\mathbf{\Theta}_k}) \cdot \delta(\overrightarrow{\mathbf{\Phi}}-\overrightarrow{\mathbf{\Phi}_k})
\end{split}
\end{equation}

\noindent where $a_k$, $\theta_k$, and $\tau_k$ are the amplitude, phase, and propagation delay of the $k$\textsuperscript{th} multipath component, $\overrightarrow{\mathbf{\Theta}_k}=(\theta_{TX},\phi_{TX})$ and $\overrightarrow{\mathbf{\Phi}_k}=(\theta_{RX},\phi_{RX})$ are the vectors of azimuth and elevation AOD and AOA of the $k$\textsuperscript{th} multipath component; $N$ is the total number of resolvable multipath components, and $\delta()$ is the Dirac delta function. Thus, each multipath tap in the tap delay line model shown in~(\ref{eq1}) has seven associated multipath parameters. Measurement-based statistical distributions for $|a_k|^2$, $\tau_k$, $\overrightarrow{\mathbf{\Theta}_k}$ and $\overrightarrow{\mathbf{\Phi}_k}$ in~(\ref{eq1}) have been extracted from 28 GHz ultrawideband propagation measurements using the \textit{time cluster - spatial lobe} (TCSL) clustering approach~\cite{Samimi15_3}, and the phases $\theta_k$ can be assumed independently and identically distributed uniformly between 0 and $2\pi$.

While~(\ref{eq1}) applies for an omnidirectional transmitter and receiver, it must be modified to reflect directional beam positioning during a real-time mmWave base-to-mobile communication link, similar to the measurements presented here. The directional CIR $h_{dir}(t)$, with fixed TX and RX antenna beam pointing directions, and for arbitrary TX and RX antenna patterns, can be expressed as~\cite{Samimi15},
\begin{equation}\label{eq2}
\begin{split}
h_{dir}(t,\overrightarrow{\mathbf{\Theta}_0},\overrightarrow{\mathbf{\Phi}_0}) &= \sum_{k=1}^M a_k e^{j\theta_k} \delta(t - \tau_k) \\
& \cdot g_{TX}(\overrightarrow{\mathbf{\Theta}_0}-\overrightarrow{\mathbf{\Theta}_k}) \cdot g_{RX}(\overrightarrow{\mathbf{\Phi}_0}-\overrightarrow{\mathbf{\Phi}_k})
\end{split}
\end{equation}

\noindent where $(\overrightarrow{\mathbf{\Theta}_0},\overrightarrow{\mathbf{\Phi}_0})$ are the fixed TX and RX beam pointing angles during the measurements, $M$ is the total number of resolvable multipath components for the $(\overrightarrow{\mathbf{\Theta}_0},\overrightarrow{\mathbf{\Phi}_0})$ pointing direction (corresponding to a subset of the total number $N$ multipath components for omnidirectional transmissions and receptions from~(\ref{eq1})~\cite{Sun15}), and $g_{TX}(\overrightarrow{\mathbf{\Theta}})$ and $g_{RX}(\overrightarrow{\mathbf{\Phi}})$ are the 3-dimensional (3-D) (azimuth and elevation) TX and RX complex amplitude antenna patterns of arbitrary multi-element antenna arrays. Here, using directional horn antennas, $10 \cdot \log_{10}(|g_{TX}(0,0)|^2) = 10 \cdot \log_{10}(|g_{RX}(0,0)|^2) = 15$ dBi.

The statistics of the path gain amplitudes $a_k$ are studied for sub-wavelength receiver motion in a realistic future mmWave scenario, where the base station and mobile terminal are capable of beamforming in very narrow azimuth and elevation pointing directions, as shown in~(\ref{eq2}).

\subsection{Measured Power Impulse Reponses}

\begin{figure}
    \centering
 \includegraphics[width=3.5in]{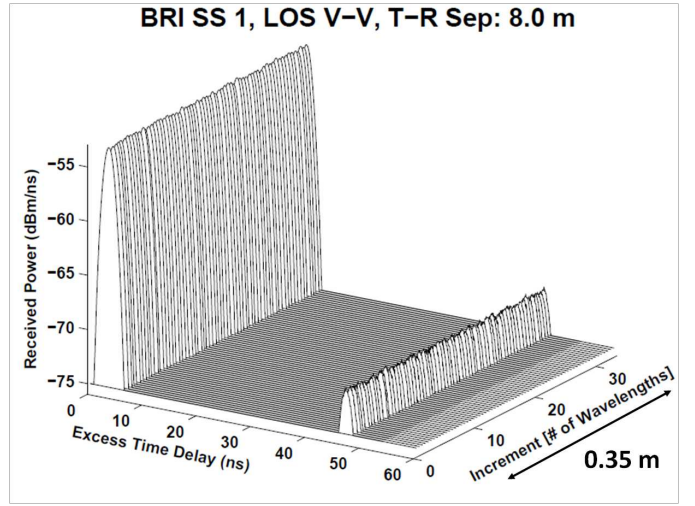}
    \caption{28 GHz small-fading PDP track measurements obtained at half-wavelength increments with a pair of 15 dBi gain directional horn antennas in LOS for a \text{V-V} antenna polarization configuration over a 33-wavelength linear track.}
    \label{fig:T1}
\end{figure}

\begin{figure}
    \centering
 \includegraphics[width=3.5in]{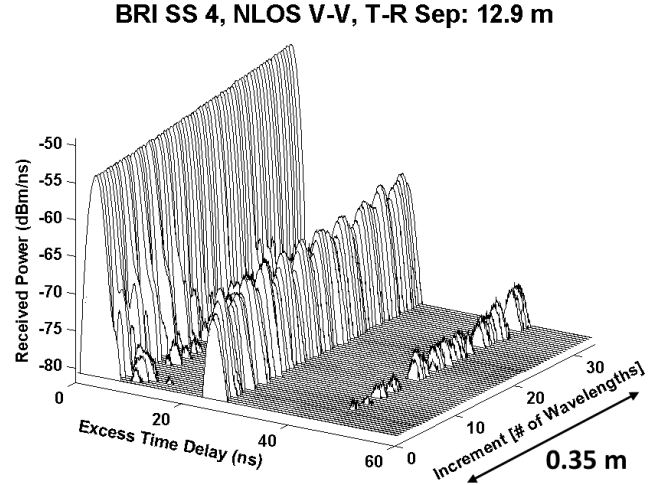}
    \caption{28 GHz small-fading PDP track measurements obtained at half-wavelength increments with a pair of 15 dBi gain directional horn antennas in NLOS for a \text{V-V} antenna polarization configuration over a 33-wavelength linear track.}
    \label{fig:T2}
\end{figure}

Figs.~\ref{fig:T1} and~\ref{fig:T2} present two 3-D plots of small-scale LOS and NLOS PDP track measurements obtained over the 33-wavelength track distance at SS 1 and SS 4, respectively, where the RX antenna was incremented over the local area towards the base station in a V-V antenna polarization configuration (refer to map in Fig.~\ref{fig:map}). In Fig.~\ref{fig:T1}, the TX and RX antennas were aligned on boresight, with a -14$^{\circ}$ TX downtilt and $+19^{\circ}$ RX uptilt with respect to horizon. The individual multipath power amplitudes are relatively constant along the spatial dimension.

In Fig.~\ref{fig:T2}, the TX antenna was downtilted by $-13^{\circ}$ and pointed towards the edge of the building corner (refer to map in Fig.~\ref{fig:map}), while the RX antenna was pointing towards the building edge corner, with a $9^{\circ}$ uptilt. Three strong multipath components were detected, where the first path propagated via diffraction around the building corner, exhibiting small amplitude fluctuations over the local area, indicating little fading. However, the second and third multipath component amplitudes, at 27 ns and 47 ns, respectively, fluctuated more significantly over the track length, resulting from the coherent sum of different multipath components arriving within the system pulse time resolution.

\subsection{Analysis of Small-Scale Fading of Path Amplitudes}

Small-scale spatial fading describes the random fluctuations of amplitudes of individual multipath components, as a mobile travels over a few wavelengths and experiences constructive interference from signals arriving within the measurement system resolution~\cite{Rap02}.

The small-scale spatial fading distributions were obtained for all individual multipath components over the local area, by discretizing the excess delay axis into bins, equal in time width to the transmitted pulse resolution, e.g., 2.5 ns. For a given delay bin, the individual multipath powers $\{|a_k|^2\}_{k=1}^{k=66}$, whose powers were greater than the noise floor, were normalized by the mean (over the spatial dimension) bin power $\overline{|a_k|^2}$ (in the linear scale). Under the assumption that small-scale fading is delay independent~\cite{Saleh87}, the small-scale fading values of all resolvable paths at all delays were grouped into one dataset. 

Amplitude distributions about the local mean were extracted over the two orthogonal measured directions for the NLOS location (SS 4),  the LOS-to-NLOS location (SS 3), and from the two LOS measured locations (SS 1 and SS 2). These were compared to a Rayleigh, Rician, and lognormal distribution. The Rayleigh distribution characterizes a scenario where many arriving multipath components have comparable delays and amplitudes, and its probability density function (PDF) is given by~\cite{Rap02},
\begin{equation}
p_X(x) = \frac{x}{\sigma_n^2}e^{-x^2/2 \sigma_n^2}
\end{equation}

\noindent where $\sigma_n$ is the standard deviation of the scattered multipath amplitudes. A Rician distribution indicates the presence of a specular dominant component in the channel over other very weak paths, and its PDF is expressed as,
\begin{equation}
p_X (x) = \frac{x}{\sigma_n^2}e^{-\frac{x^2+A^2}{2\sigma_n^2}} I_0 \bigg(  \frac{Ax}{\sigma_n^2} \bigg)
\end{equation}
\begin{equation}
K = \frac{A^2}{2\sigma_n^2}
\end{equation}

\noindent where $I_0()$ is the modified Bessel function of the first kind and zero order, $A$ is the amplitude of the dominant path, and $\sigma_n$ is the standard deviation of all other weak path amplitudes. The PDF of a lognormal distribution is given by~(\ref{LOG}),
\begin{equation}\label{LOG}
p_X(x) = \frac{1}{\sqrt{2 \pi}\sigma x}e^{-(\log x - \overline{x})^2/2\sigma^2}
\end{equation}
\noindent where $\overline{x}$ and $\sigma$ are the mean and standard deviation, respectively.
 
\begin{figure}
    \centering
 \includegraphics[width=3.5in]{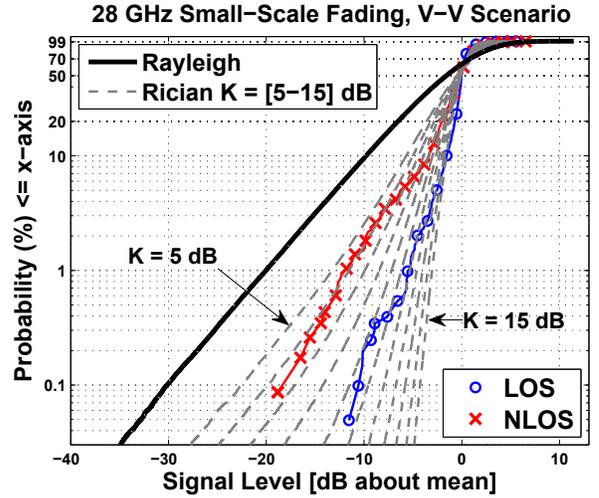}
    \caption{CDFs of the 28 GHz small-scale fading of individual path (voltage) amplitudes in LOS and NLOS, for the co-polarized V-V scenario. Rayleigh and Rician distributions are also plotted for various values of $K$-factors ranging from 5 dB to 15 dB, in 1 dB increments.}
    \label{fig:T3}
\end{figure}

\begin{figure}
    \centering
 \includegraphics[width=3.5in]{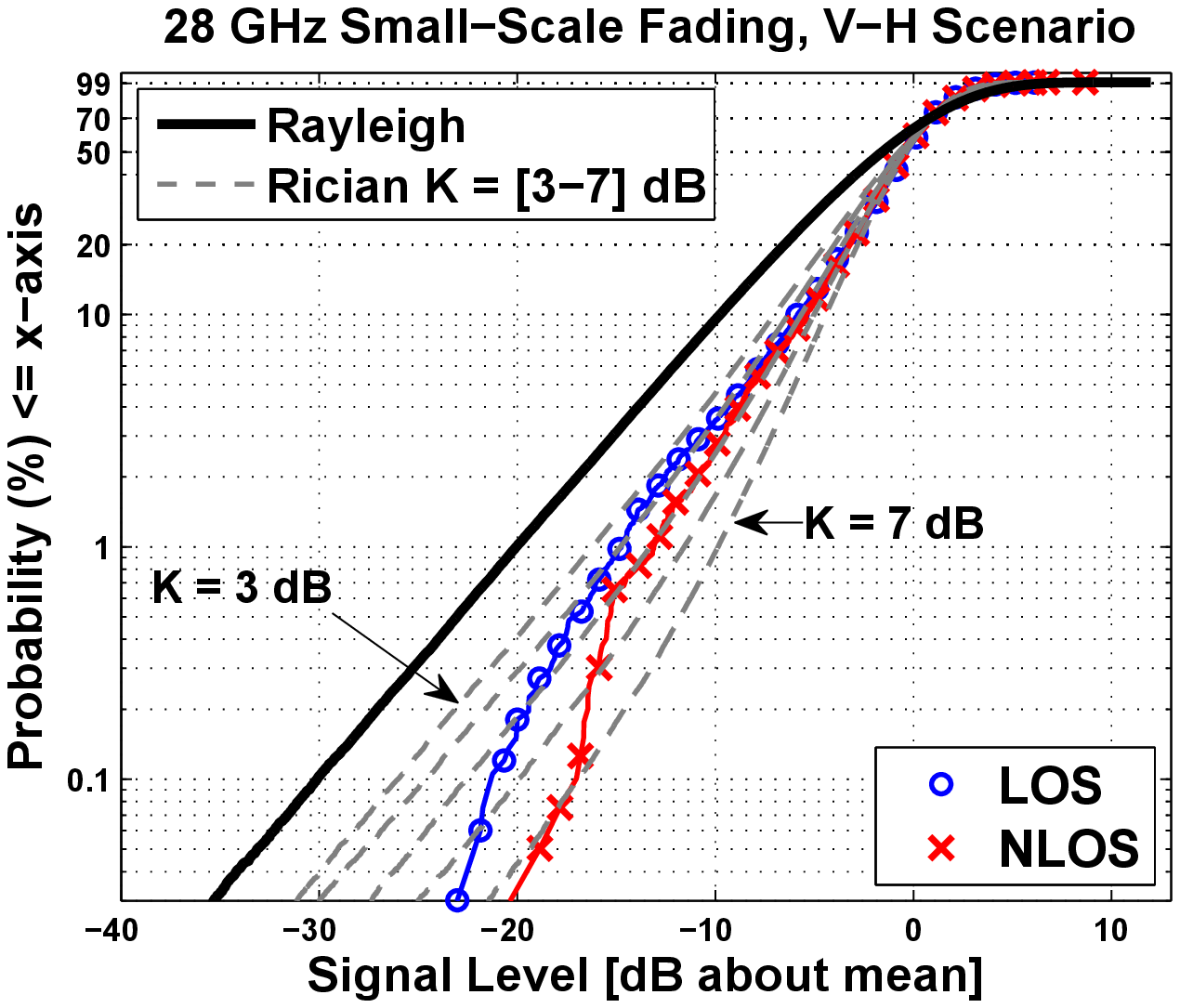}
    \caption{CDFs of 28 GHz small-scale fading of individual path (voltage) amplitudes in LOS and NLOS, for the cross-polarized V-H scenario. Rayleigh and Rician distributions are also plotted for various values of $K$-factors ranging from 3 dB to 7 dB, in 1 dB increments.}
    \label{fig:T4}
\end{figure}

\begin{figure}
    \centering
 \includegraphics[width=3.5in]{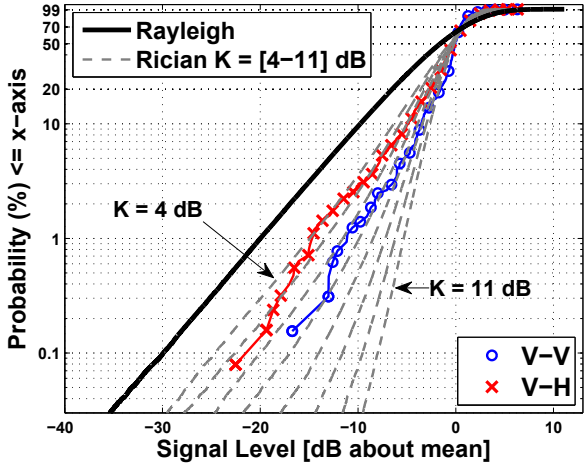}
    \caption{CDFs of 28 GHz small-scale fading of individual path (voltage) amplitudes in a transitional LOS-to-NLOS scenario, in co- and cross-polarization scenarios. Rayleigh and Rician distributions are plotted for various values of $K$-factors ranging from 4 dB to 11 dB, in 1 dB increments.}
    \label{fig:T5}
\end{figure}

The empirical cumulative distribution functions (CDFs) for $|a_k|^2/\overline{|a_k^2|}$ in V-V co-polarized and V-H cross-polarized scenarios in LOS, NLOS, and LOS-to-NLOS are shown in Figs.~\ref{fig:T3}, ~\ref{fig:T4}, and ~\ref{fig:T5} with the CDF of a Rayleigh distribution, and Rician distributions plotted for various values of $K$-factors in increments of 1 dB. The Rician distribution provided the best fit to the data, over the Rayleigh and lognormal distributions. For the V-V co-polarized data (Fig.~\ref{fig:T3}), the curves are bounded by two Rician distributions with $K$-factors of 9 dB and 15 dB in LOS, and 5 dB and 8 dB in NLOS. For the V-H cross-polarized data (Fig.~\ref{fig:T4}), both the LOS and NLOS CDFs are bounded by two Rician distributions with $K$-factors of 3 dB and 7 dB. Similarly, in the LOS-to-NLOS scenario, the empirical CDFs lie between two Rician distributions, whose $K$-factors are 4 dB and 6 dB in the V-V scenario, and 6 dB and 10 dB in the cross-polarized V-H scenario.

Note that the Rayleigh distribution underestimates the measurement data, indicating that for large enough signal bandwidth (here 800 MHz RF null-to-null), the measurement system is able to resolve individual, or the coherent sum of just a few, multipath components arriving within the path resolution. Table~\ref{tbl:T1} summarizes the ranges of Rician $K$-factors as a function of polarization scenarios and environment types obtained from the 28 GHz measurements.

\begin{table}
\centering
\caption{Summary of $K$-factors for the Rician distributions, that describe the path (voltage) gains $a_k$ in~(\ref{eq1}), obtained from 28 GHz directional small-scale fading measurements over a local area in different environments, for V-V and V-H polarization configurations.}

\begin{tabular}{|c|c|c|}

\hline
\textbf{Environment}	& \bm{$K_{VV}} \textbf{ [dB]}$	& \bm{$K_{VH}} \textbf{ [dB]}$ \\ \hline

LOS				& 9 - 15					& 3 - 7				\\ \hline
NLOS				& 5 - 8					& 3 - 7				\\ \hline
LOS-to-NLOS			& 4 - 7 					& 6 - 10				\\ \hline
\end{tabular}
\label{tbl:T1}
\end{table}

\subsection{Spatial Autocorrelation of Individual Multipath Amplitudes}

\begin{figure}
    \centering
 \includegraphics[width=3.5in]{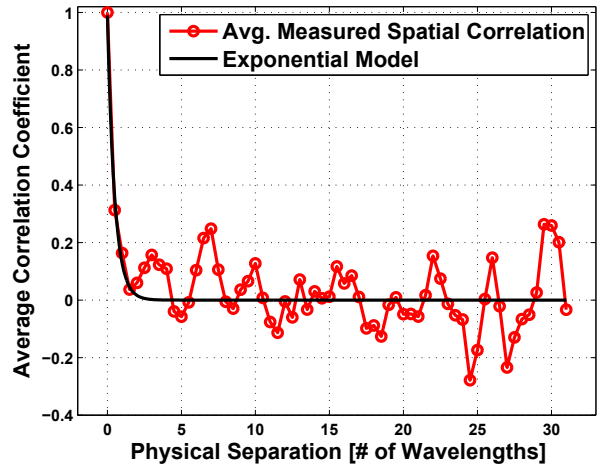}
    \caption{28 GHz empirical spatial autocorrelation function at the RX for the LOS V-V scenario, averaged over all possible excess delay bins. The model~(\ref{eq3}) is shown, where the model parameters (A, B, C) were obtained using the MMSE method.}
    \label{fig:T6}
\end{figure}

\begin{figure}
    \centering
 \includegraphics[width=3.5in]{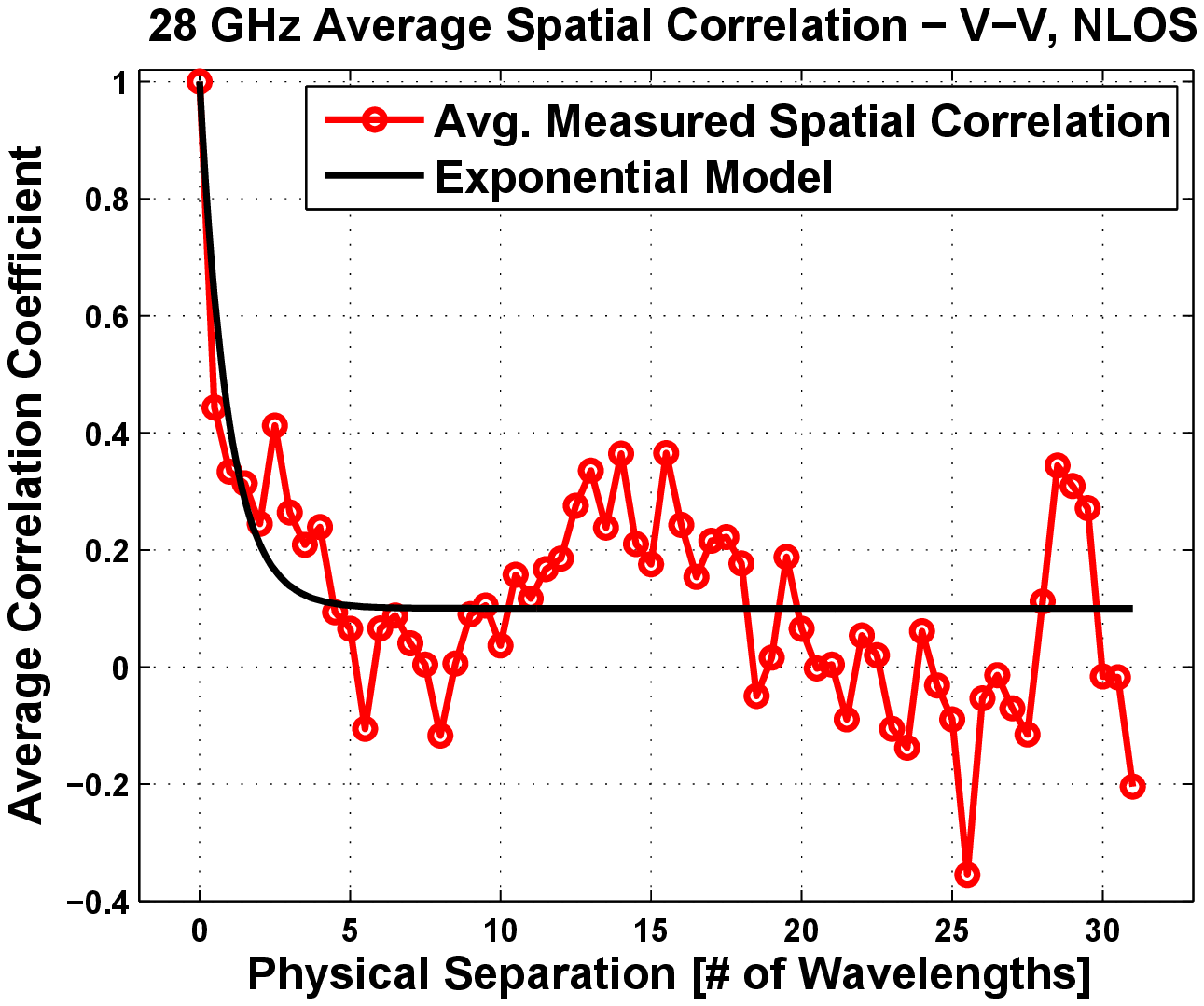}
    \caption{28 GHz empirical spatial autocorrelation function at the RX for the NLOS V-V scenario, averaged over all possible excess delay bins. The model~(\ref{eq3}) is shown, where the model parameters (A, B, C) were obtained using the MMSE method.}
    \label{fig:T7}
\end{figure}

The spatial autocorrelation indicates the level of similarity in multipath signal amplitudes over fractions of wavelengths, and has been used to recreate realistic spatial correlations in multi-element antenna simulations for mmWave system design~\cite{Samimi16}. The average spatial autocorrelation coefficients were computed from~(\ref{Eq1}) for the LOS, NLOS, and LOS-to-NLOS environments in both co- and cross-polarized antenna configurations, where E[] is the average over RX location and environment (i.e., two orthogonal measured directions for specified environment), $\Delta X$ is the physical separation between two adjacent track positions, and is equal to integer multiple of $\frac{\lambda}{2} = 5.35$ mm, $A_K(T_K,X_l)$ is the multipath amplitude power at track position $l$ and at bin delay $K$. Fig.~\ref{fig:T6} and Fig.~\ref{fig:T7} show typical measured average spatial autocorrelation functions of individual multipath amplitudes at the RX in LOS and NLOS environments for the V-V co-polarized scenario, averaged over all possible excess delay bins. The empirical curves were fit to an exponential model of the form~\cite{Karttunen98},
\begin{equation}\label{eq3}
f(\Delta X) = A e^{-B \Delta X}-C
\end{equation}

\noindent where $A$, $B$, and $C$ are constants that were determined using the minimum mean square error (MMSE) method, by minimizing the error between the empirical curve and theoretical exponential model shown in~(\ref{eq3}). In Fig.~\ref{fig:T6} and Fig.~\ref{fig:T7}, the constants were determined to be $A=0.99$, $B=2.05$, $C=0$, and $A=0.9$, $B=1.05$, $C=-0.1$, respectively. Table~\ref{tbl:T2} summarizes the model coefficients as a function of polarization and environment type.

\begin{figure*}
\begin{equation}\label{Eq1}
\rho(i \Delta X) = \frac{E \big[ \big( A_K (T_K,X_l)- \overline{A_K(T_K,X_l)} \big)    \big( A_K (T_K,X_l+i\Delta X)- \overline{A_K(T_K,X_l +i \Delta X )} \big)       \big] }{\sqrt{E \bigg[ \big( A_K (T_K,X_l)- \overline{A_K(T_K,X_l)} \big)^2  \bigg] E\bigg[ \big( A_K (T_K,X_l+ i \Delta X)- \overline{A_K(T_K,X_l + i \Delta X )} \big)^2       \bigg]}}, i = 0, 1, 2, ...
\end{equation}
\hrulefill
\vspace*{4pt}
\end{figure*}

\begin{table}
\centering
\caption{Summary of model parameters (A, B, C) in~(\ref{eq3}) obtained using the MMSE method, to estimate the empirical spatial autocorrelation functions.}

\begin{tabular}{|c|c|c|}

\hline
\textbf{(A, B, C)}		& \textbf{V-V}				& \textbf{V-H}				 \\ \hline

LOS				& (0.99, 2.05, 0)				& (1.0, 0.9, 0.05) 				\\ \hline
NLOS				& (0.9, 1.05, -0.1)				& (1.0, 1.9, 0)				\\ \hline
LOS-to-NLOS			& (0.9, 1.9, -0.3)				& (0.9, 1.05, 0)				\\ \hline
\end{tabular}
\label{tbl:T2}
\end{table}

\section{Conclusion}

This paper presented 28 GHz ultrawideband outdoor small-scale fading measurements performed in LOS and NLOS environments over a local area. The voltage path gain amplitude, under narrow TX and RX beam pointing positioning, was shown to be best characterized by a Rician distribution, with $K$-factors ranging from 9 - 15 dB and 5 - 8 dB in LOS and NLOS V-V scenarios, respectively, and 3 - 7 dB in both LOS and NLOS V-H scenarios. For very wideband mmWave channels, the presented data indicates that path amplitudes are no longer Rayleigh-distributed in NLOS environments, suggesting that individual, or the coherent sum of just a few, multipath components are sufficiently resolved by the measurement system. Average spatial autocorrelation functions of individual multipath components were computed, showing that signals typically reach zero correlation after 2 and 5 wavelengths in LOS and NLOS environments, respectively. With the provided models, the small-scale fading statistics of wideband multipath amplitudes can be simulated over a local area in the study of multi-element antenna diversity schemes for next generation mmWave systems, such as in~\cite{Samimi16}.

%

\bibliographystyle{IEEEtran}
\bibliography{bibliography_MSThesis}
\end{document}